\documentclass[preprint]{vgtc}                          
\ifpdf
  \pdfoutput=1\relax                   
  \usepackage{graphicx}                
  \DeclareGraphicsExtensions{.pdf,.png,.jpg,.jpeg} 
\else
  \usepackage{graphicx}                
  \DeclareGraphicsExtensions{.eps}     
\fi%

\graphicspath{{figures/}{pictures/}{images/}{./}} 
\usepackage{subfig}
\usepackage{framed}
\usepackage{microtype}                 
\PassOptionsToPackage{warn}{textcomp}  
\usepackage{textcomp}                  
\usepackage{mathptmx}                  
\usepackage{times}                     
\usepackage{cite}                      
\usepackage{tabu}                      
\usepackage{booktabs}                  

\vgtccategory{Research}

\vgtcinsertpkg

\title{Cube2Pipes : Investigating Hybrid Gameplay Using AR and a Tangible 3D Puzzle}

\author{Sukanya Bhattacharjee\\ %
\scriptsize
Department of Computer Science and Engineering\\
\scriptsize IIT Bombay\\
\scriptsize sukanyabhat@cse.iitb.ac.in
\and Parag Chaudhuri\\ %
\scriptsize
Department of Computer Science and Engineering\\
\scriptsize IIT Bombay\\
\scriptsize paragc@cse.iitb.ac.in
} %

\abstract{We present our game, Cube2Pipes, as an attempt to investigate a unique gameplay design where we use a tangible 3D spatial puzzle, in the form of a $2\times2$ Rubik's Cube, as an interface to a tabletop mobile augmented reality (AR) game. The game interface adapts to user movement and interaction with both virtual and tangible elements via computer vision based tracking. This game can be seen as an instance of generic interactive \emph{hybrid} systems as it involves interaction with both virtual and real, tangible elements. We present a thorough user evaluation about various aspects of the gameplay in order to answer the question as to whether hybrid gameplay involving both real and virtual interfaces and elements is more captivating and preferred by users, than standard (baseline) gameplay with only virtual elements. We use multiple industry standard user study questionnaires to try and answer this question. We also try to determine whether the game facilitates understanding of the spatial moves required to solve a Rubik's Cube, and the efficacy of a tangible puzzle interface to a tabletop AR game.} 

\CCScatlist{
  \CCScatTwelve{Human computer interactions}{Mixed/Augmented Reality}{Hybrid games}{Tangible 3D puzzle interface}
}

\begin{document}

\teaser{
 \includegraphics[width=\textwidth]{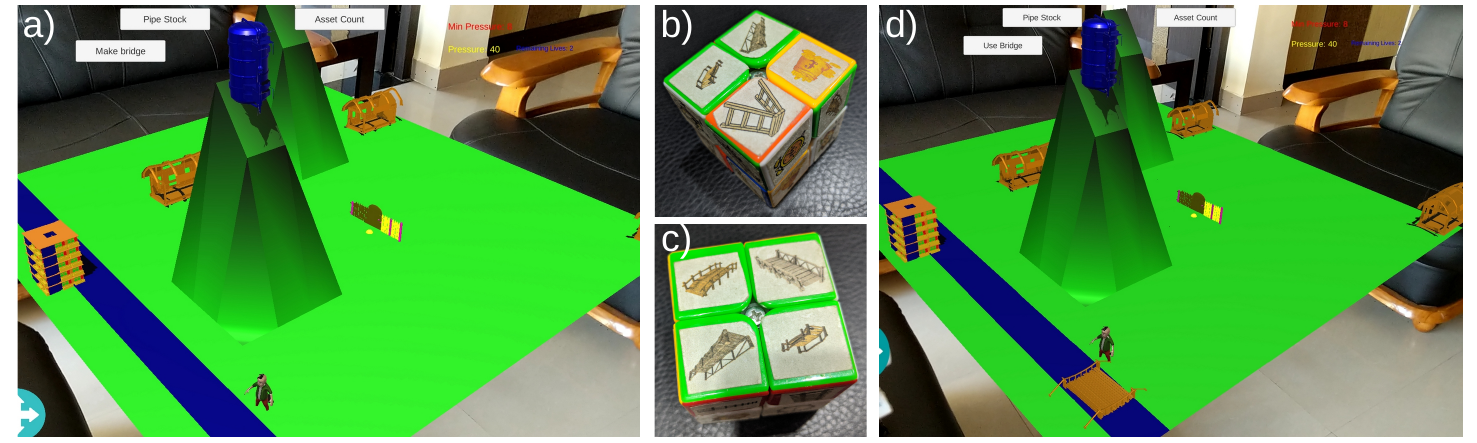}
 \caption{At some instant, to progress in the Cube2Pipes game (a) the player needs a bridge asset to cross the river. (b) A Rubik's Cube puzzle needs to be solved to get assets in the game. (c) A layer of the cube is solved to get the bridge asset. (d) The player gets the bridge asset and the game proceeds.}
 \label{figteaserImage}}
 
\maketitle

\section{Introduction}\label{intro}
Augmented Reality (AR) tries to enhance our view of the real world by placing virtual elements in it. This has led to its use in many applications ranging from retail to entertainment. Gaming is an area that has seen many interesting uses of AR. Games like Pokemon Go~\cite{pokemonGo} engage the user by letting them interact with virtual elements in the real world. More recent games like Minecraft Earth~\cite{minecraft} extend this experience by adding persistence and shared gameplay. 
\emph{Hybrid}~\cite{kankainen2019} games use both virtual and real, tangible elements for gameplay. In this paper, we develop and explore a mobile hybrid game called \textit{Cube2Pipes}. The game is played on a virtual game board that can be overlaid on any physical flat surface using AR. In order to progress in the game, the player has to collect assets. Assets are obtained by the player by solving a real, tangible 3D spatial puzzle in the form of a $2\times2$ Rubik's Cube~\cite{rubik}. 

The primary research interest of the paper is to investigate if this kind of hybrid gameplay, using virtual and tangible elements, is more interesting and appealing to a user than games that involve only virtual elements. In order to establish this we present a thorough user evaluation of the interfaces and the interactions in the game. We would like to present this game as an example of the kind of tasks that can benefit from hybrid gameplay. It can be considered as an instance of a class of tasks that will benefit from such interactions. The questions asked in the user study try to ascertain the ease-of-use and involvement of generic hybrid gameplay so that the inferences can hold for the entire class of tasks we aim to represent.

We present various aspects for the design of our game in subsequent sections. It is not an uncommon practice in video game design to use one challenge in facilitating the progress of another challenge. This not only makes the game more interesting but also increases the involvement of the user by adding another level of cognitive load. While in a traditional video game these challenges exist completely in the virtual world, we use a tangible, real $2 \times 2$ Rubik's Cube puzzle as the challenge mechanism to interact with the game. Thus, Cube2Pipes not only leverages the freedom that mobile AR offers in navigating a 3D virtual world, it also involves both visual and tactile senses of the user by this interesting hybrid gameplay. We present a user study to probe into our claims about the potential of such a system in engrossing a user in the gaming experience, and to show that the users are able to easily understand the spatial moves required to solve a Rubik's Cube during the gameplay. An important aspect of the game interface design is to respond to user actions. This involves adapting to user's movement around the virtual, planar game board that is ensured by using robust computer vision aided tracking. It also involves tracking the user interaction with the Rubik's cube so that the users can be assisted with augmented visual cues, in their quest to get the required assets. Our paper tackles these design challenges and attempts to answer the following research questions:

-- \textbf{R1:} \emph{Are hybrid gameplay designs, like Cube2Pipes, more interesting and captivating for a user, than virtual only game designs?}

-- \textbf{R2:} \emph{Do the augmented visual cues on the Rubik's Cube assist the user in deciphering the complex spatial moves required to solve it?}

-- \textbf{R3:} \emph{Does the use of a tangible 3D puzzle interface create too much cognitive load for the user?}

We explore related literature in Section~\ref{relWork}. Then we describe the design of our game in Section~\ref{gameMethod} and provide relevant details about the responsive interface elements in our game. Our detailed user evaluation is presented in Sections~\ref{userEval},~\ref{gameMechEval} and~\ref{rubikCubeEval}. We briefly describe the performance of the game in Section \ref{otherPar} and finally conclude with a discussion on inferences we draw from it in Section~\ref{disc}.
\section{Related work}\label{relWork}
Many works in existing literature explore the game play and game design in AR. In an early work, Lee et al.~\cite{lee2005tarboard} propose an AR based game set up for making traditional board games and card games more interactive. The physical game elements are placed on top of a glass table which has one of the cameras fitted at its bottom. This camera uses the reflection of the marker on the mirror to track it. The other camera is used for augmenting virtual content on top of the marker with the help of the first camera. Huynh et al.~\cite{huynh2009art} propose a handheld AR based board game with tangible game elements. Hexagonal markers and different game tokens are used as game elements. The markers are used for creating the maps which form the base on which the virtual towers appear. With an elaborate user study, they are able to conclude that such an AR interface is able to ensure social interaction in a collaborative set-up and sense of presence due to the tangible components although more stable augmentations and improvement to the interface are desirable. A method to detect the physical elements of a board game to augment it with the virtual objects is presented in~\cite{molla2010augmented}. They use feature point matching for locating the board and placing 3D boxes around the physical elements. The detection and color recognition locates the elements on which the virtual objects can be placed and oriented accordingly. Hagbi et al.~\cite{hagbi2010place} devise a method to use sketch based annotations to design a game in mobile AR, augmented on any planar real object like a napkin or a whiteboard. Patricio et al.~\cite{patricio2018solarsystemgo} present a mobile AR based game designed primarily to make children aware about the solar system. The environment geometry and lighting is interpreted while augmenting the real world with a virtual solar system. Recently, Rematas et al.~\cite{rematas2018soccer} leveraged deep neural networks to visualize soccer game videos in 3D using an AR platform. Though this work is not focused on the 3D game designing, but provides a way to bring recorded game telecasts to the 3D real world. An elaborate discussion on AR based games and their various aspects is provided by Li et al.~\cite{li2017augmented}.

Many commercial games like Pokemon Go~\cite{pokemonGo} and Minecraft Earth~\cite{minecraft} take advantage of AR for gameplay. Pokemon Go makes the user move in the real world and uses location awareness to place virtual pokemons in their surroundings. Minecraft Earth lets the user build block structures and place them in the real world. Additionally, users can share and explore this structure with their friends. TiltFive~\cite{tiltfive} has recently introduced a number of tabletop games which can be played in the AR environment using their Holographic gaming system comprising of either a mobile phone or a pair of AR glasses, a handheld device, a set of cards and a game-board. LegoAR~\cite{legoAR} presents a series of popular Lego games blended with the AR environment where the physical Lego toys are used for interacting with the virtual scene in the game viewed using a mobile device. Bergig et al.~\cite{bergig2011out} design a Rubik's Cube based game to provide resources to a set of $6$ villages augmented to each face of the cube. A sticker kit is used to track the movement of cube. The gameplay follows the Rubik's Cube mechanics to achieve the end goal, where the player always checks all the faces to ensure that any shuffle does not disturb the previous arrangement that can damage resources of the villages. Games like MergeCube~\cite{mergecube} and CubeAR~\cite{cubeAR} use a cube as the physical premise for the augmentation.  MergeCube uses a customized cube with special markers pasted on each of its faces. In MergeCube, different virtual scenes are augmented depending on the marker and the augmentation is viewed using a mobile device. CubeAR attempts to guide the user to solve a Rubik's Cube by augmenting arrows in the direction of required spatial moves. It scans the entire cube as a one time step and then overlays the arrow according to the current visible face. 

While the above mentioned works have explored augmentation of board games and other physical space in innovative ways, works like \cite{de2017modelling,de2016relational, kankainen2019hybrid, kankainen2014understanding, kankainen2016interplay, kosa2018tabletop} study the hybrid systems and game designs in multiple dimensions. To emphasise, by hybrid gameplay we mean a game design where both virtual and tangible components are part of the game interface. Based upon the literature and case studies, \cite{de2016relational} describes a relational model for hybrid games which is capable of showing the relations between the entities like player, environment and playable objects helping in considering various scenarios and characteristics of such systems. In \cite{de2017modelling}, de et al. update the relational model of \cite{de2016relational} by studying the feedback from a set of students in a classroom setup who were taught about the hybrid system designing using the relational model. Kankainen et al. propose a classification of the role of smart devices in the premise of hybrid games in \cite{kankainen2014understanding}. With an intention to further define design goals for hybrid games, Kankainen et al. \cite{kankainen2016interplay} attempt to study the difference in the impact of the physical and digital versions of a game on the user depending on how the user trade offs between various factors of the design. Kosa et al. \cite{kosa2018tabletop} study various responses of users available on internet about hybrid games and try to create a taxonomy of both positive and negative responses along with their reasons. In \cite{kankainen2019hybrid}, Kankainen et al. put across many guidelines in terms of the design choices based upon extensive investigations into hybrid systems. Works like \cite{gunther2018checkmate, narazani2019extending, tyni2013dimensions} are among many such works which have studied some particular hybrid game designs with an objective of deciphering their usability in various dimensions. The work in this paper falls into a similar category with our goals of being able to unfold multiple dimensions of hybrid games, in particular their cognitive load and sense of presence, and their potential usage in assistance for solving some puzzle-like problem while maintaining its usability.


In particular, our contributions include:
\begin{itemize}
    \item We present a novel \emph{hybrid} gameplay design that uses a real, tangible 3D spatial puzzle or challenge interface to a tabletop AR game.
    \item We present a thorough user evaluation of the proposed gameplay and all associated interfaces with respect to their ease-of-use and effectiveness in captivating users, which can potentially provide interesting insights about generic hybrid interactive systems.
\end{itemize}

\section{Game Design}\label{gameMethod}
\begin{figure}[h!tb]
    \centering
    \begin{minipage}{\columnwidth}
        \centering
        \subfloat[Visual cue for rotating a layer.\label{figaugCube_a}]{\includegraphics[width=0.5\linewidth]{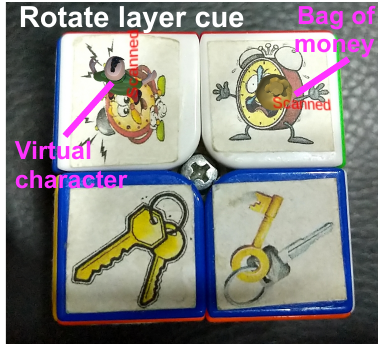}}~~~~%
        \subfloat[Visual cue for turning the entire cube over.\label{figaugCube_b}]{\includegraphics[width=0.5\linewidth]{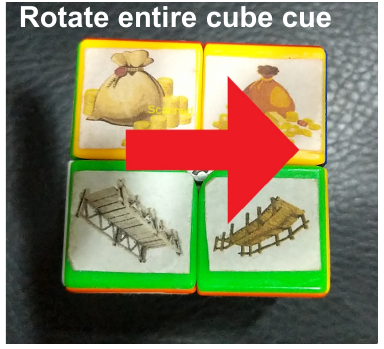}}
        \caption{Augmented visual cues on the cube.}
        \label{figaugCube}
    \end{minipage}
\end{figure}
We want to involve the user in solving a challenging tangible puzzle, which is required to proceed ahead in completion of a cognitive task within the premise of an AR based game. We want to do this to uncover various merits and demerits of such a hybrid AR based gameplay. As an instance of the cognitive task, the objective of our game Cube2Pipes is that the user has to lay pipes between a water tank and multiple houses placed on a virtual terrain that represents a small town. In order to navigate the terrain and complete the task, the player has to acquire various assets like bridges, ladders, pipes, etc. The assets are obtained by solving various stages of a real, tangible $2\times2$ Rubik's Cube and this is setup in a way that gathering all the assets solves the cube. The directions to solve the cube are computed by the game engine and overlaid on the cube appropriately in form of visual cues (see Figure~\ref{figaugCube}). We choose such a tangible interaction in this game with the objective of unfolding the potential of visual augmentations in helping understand solutions to popular spatial puzzles along with its effectiveness for elevating interest of the users in a relatively simple virtual game task such as the pipe laying task.

Solving the Rubik's cube is an instance of gameplay involving a tangible component.
We assign the same semantic meaning to the similarly colored cells of the Rubik's Cube by using similar markers for them. This can also be considered as an instance of different configuration designs possible for a Rubik's cube. The user is given the flexibility to choose the way she wants to play the game. Novice users who have little or no experience with a Rubik's cube can use the visual cues we present to solve an unsolved or semi-solved cube and play the game, while users who already know how to solve a Rubik's Cube can scan a solved cube and proceed. In order to balance the overall cognitive load for the user, we choose a simple task like pipe laying and a relatively more challenging task of solving a Rubik's Cube so that the users are able to focus more on the objective of the game, instead of getting lost in solving the puzzle. For evaluation purposes, we consider the purely virtual component comprising of laying pipes on the virtual terrain as a \emph{baseline}. To this we add a tangible component in the form of a Rubik's Cube, to get the complete hybrid game.

Based upon the guidelines given in \cite{kankainen2019hybrid}, our game design fits into the following criteria --- \textit{accessibility} is taken care of by presence of Rubik's cube in which basic rotation mechanism is easy to capture; \textit{added value} is ensured by adding the tangible challenge to the virtual challenge which not only expands the game features but also guides users in solving the cube; \textit{automation} is one of the key feature in our game as tasks like cube solving to placing of assets on virtual game board is guided by the game engine (more details in next section); \textit{aesthetics} is incorporated by using AR to enrich the visual experience; \textit{availability} is ensured by using the any ARCore enabled mobile phone device for playing the game; \textit{scalability} is added in terms that more game features can be added to it at any stage by the developer; \textit{customizability} is ensured by giving the user the choice to play the game with only virtual elements by scanning a fully or semi- solved cube or play with the tangible component; \textit{tangibility} is implicitly there because of presence of the cube; \textit{integration} is taken care of by making the cube solving an requirement to finish the game. Considering the characterization of the smart device given in \cite{kankainen2014understanding}, the mobile phone falls into the smart device overseeing play in our case as we use ARCore to keep track of the players' movement through various markers. In an attempt to keep up with the pointers given in \cite{kosa2018tabletop} about acceptance of augmented tabletop games, we kept our gameplay design quite simple to ensure ease of use while keeping the gameplay interesting along with providing stable augmentations such that users can move around it as they would do in case of a real tabletop game for performing some actions.

Next we discuss the  responsive user interface (UI) that we created for our game. Additional game design and implementation details may be found in the \emph{supplementary PDF document} submitted with this paper.

\subsection{Responsive UI design}\label{intelUI}
\begin{figure}[h!tb]
    \centering
    \begin{minipage}{\columnwidth}
       \centering
        \subfloat[Game board marker used to place the virtual terrain.\label{figgameMarker_a}]{\includegraphics[width=0.4\linewidth]{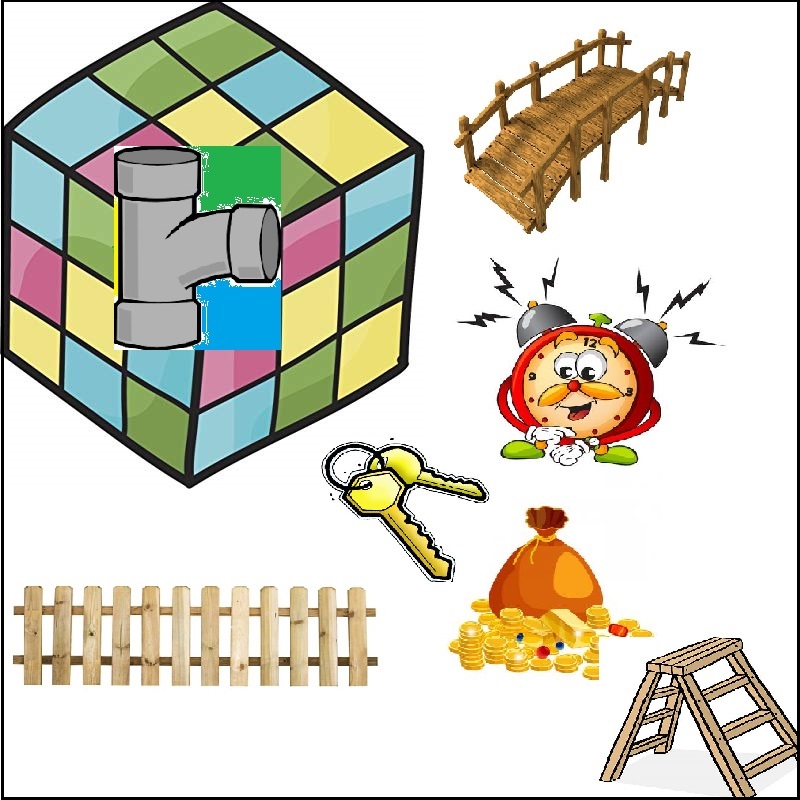}}~~\hspace{10px}%
      	\subfloat[Set of markers used on the cube.\label{figgameMarker_b}]{\includegraphics[width=0.37\linewidth]{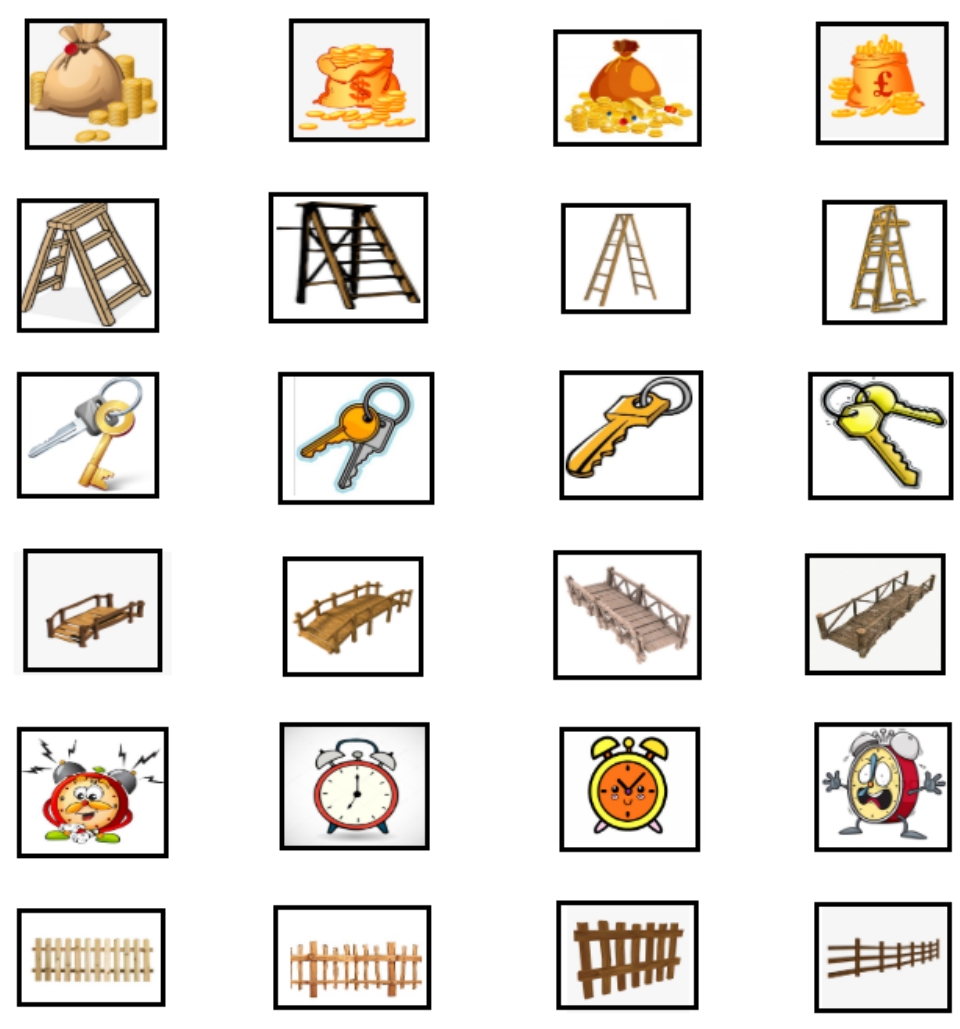}}
        \caption{Markers used to assist tracking in Cube2Pipes.}
        \label{figgameMarker}
    \end{minipage}%
\end{figure}

\begin{figure}[h!t]
    \centering
    \subfloat[Static assets on the terrain and game UI buttons.\label{figterrainAssets_a}]{\includegraphics[width=0.45\linewidth]{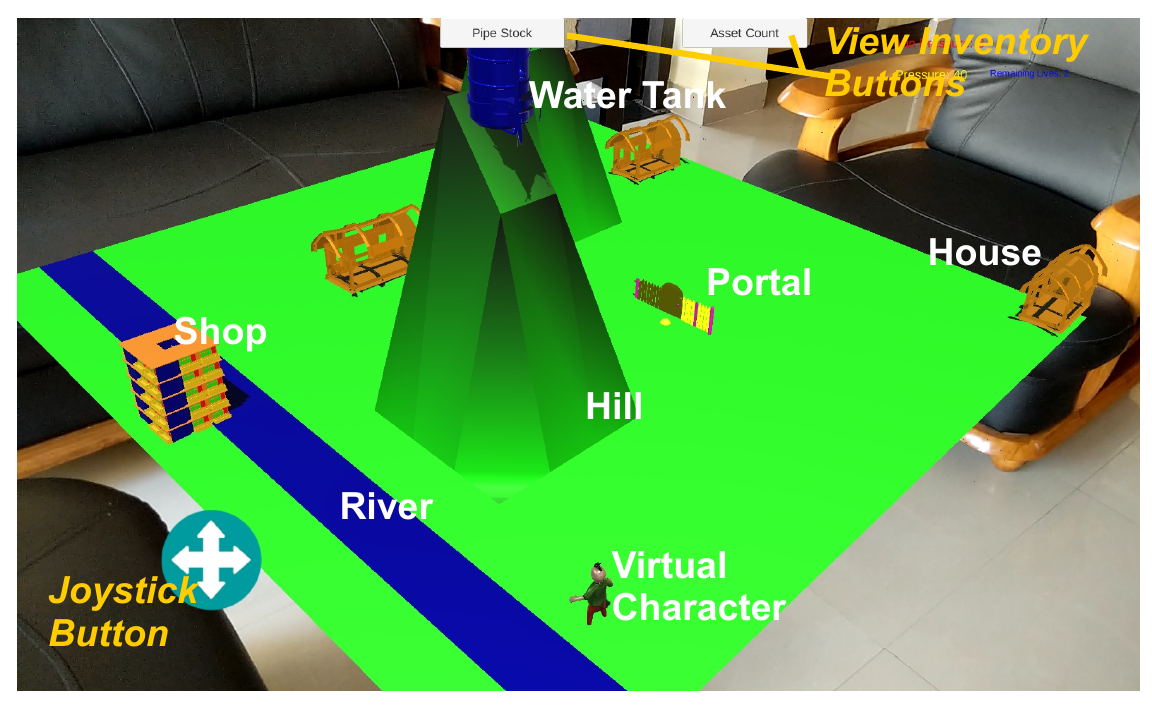}}~~~~
    \subfloat[Dynamic assets on the terrain.\label{figterrainAssets_b}]{\includegraphics[width=0.45\linewidth]{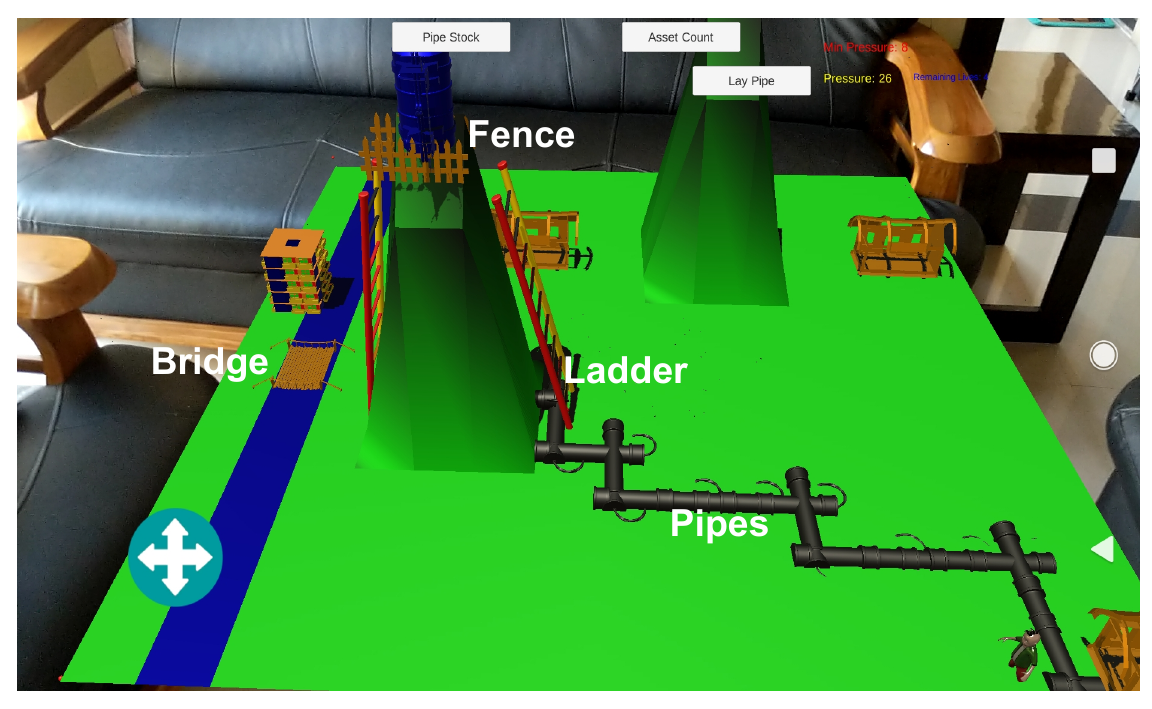}} 
    \caption{Game assets and UI. (Asset models courtesy \url{https://free3d.com} and \url{https://www.cgtrader.com}.)}
    \label{figterrainAssets}
\end{figure}

The game UI adapts in response to action taken by the user. This is accomplished via computer vision techniques that detect and track various kinds of markers by matching feature points between the marker images stored in a database and the images in the live camera feed using ARCore\cite{arcore}. We have two sets of markers, a game board marker for anchoring the virtual terrain to a plane surface, and another to overlay visual cues atop the Rubik's cube (Figure~\ref{figgameMarker}). The UI is designed in a way that the game board augmentations are in world space and the GUI elements like buttons and inventory displays corresponding to current state of the game are in screen space. This makes the game display more realistic while not cluttering the screen too much. 
Figure~\ref{figterrainAssets_a} shows an example terrain with the static assets  placed.

During gameplay, static assets are automatically placed by the game engine while other dynamic assets are interactively placed by the user (see Figure~\ref{figterrainAssets}). The game logic works by snapping elements to an underlying game grid to compute occupied and navigable regions, and its working is mostly transparent to the user. The game engine is also able to pop up GUI buttons for usage of already placed assets like ladders or bridges, avoid collisions with assets like houses and disallow placing pipes with incompatible diameter.
To tie in the tangible component of the game, a seamless transfer of gameplay is essential. The game initiates the cube solving mechanism via the usage of portals placed on the virtual terrain. When the virtual character approaches a portal, the gameplay shifts from the virtual terrain game board to the Rubik's cube. Depending upon current state of the cube face, as seen on camera feed and the required correct state, a sequence of rotations is determined by the game engine. Since the user may make a mistake in solving the cube, the game engine further incorporates a logic that can detect a single incorrect cube movement done by the user and suggest corrections. If the user performs multiple cube moves without paying heed to the cues, she is asked to restart the game. The interface is able to detect fully-solved or semi-solved cube and respond accordingly with changes to the game state. 

Thus our intuitive game UI lets the user play the game freely while minimizing common errors and providing assistance as and when required. We believe such adaptive interfaces are crucial to the success of hybrid interactive systems to lessen the cognitive load that the usage of mixed mode interfaces must entail.
\section{User Evaluation Setup}\label{userEval}
In order to perform a thorough user evaluation of the game design choices and gameplay we use a combination of different standard techniques. We choose to have a combination of these techniques to uncover different aspects of the system ranging from the creativity in the gameplay to the cognitive load of the tasks and realism of the interface. After playing the game, the user is provided with a questionnaire where she needs to give a score on the scale of $5$ ($1$ - worst/disagree to $5$ - best/agree) to each question. We post-process these scores suitably to get the required statistical measures for further analysis. We use a general questionnaire of $10$ questions to evaluate the overall gaming experience, NASA TLX~\cite{hart1988development} to evaluate the cognitive load caused by addition of the tangible element to the virtual gameplay, and Presence~\cite{witmer1998measuring} questionnaires to study the perception of presence for the user in the hybrid game. We also evaluate the effectiveness of the visual cues provided in the game for solving the Rubik's Cube to more commonly available written instructions for the same. For this, the users were asked to solve the Rubik's cube three times in this order - with our visual cues method, written instructions and on their own. After the users were satisfied with their understanding of solving the cube, they were asked to rate each of the solution methods on the scale of $5$ along with the NASA-TLX and presence questionnaires about the visual cues method.

Ten ($6$ male and $4$ females) users participated in evaluating the Rubik's Cube solution methods, eight ($4$ male and $4$ female) users volunteered to evaluate the complete game, and five ($3$ males and $2$ females) users further participated in evaluating the baseline version of the game. We considered each of the $4$ set of evaluations to be a separate entity so in total we have $28$ different measurements overall. All of them are aged between $25$-$30$ years and are familiar with smartphones and computer games. They all had little experience with AR setups. The minimum requirement for choosing the volunteers were that they had no known vision problem which might disrupt the experience and they had no knowledge of cube solving in general. The evaluation was carried out indoors and users were provided with an ARCore enabled smartphone, a $2 \times 2$ Rubik's cube with markers and the terrain marker. All of them were given around $10$-$15$ minutes to get familiar with the gameplay mechanism and then they were asked to play the game. No explicit instructions were given about the gameplay - the users learnt the interface by playing the game. In general, hybrid interaction systems can have audiences from any age group but considering the particular system we propose, the target audience is relatively younger with a certain interest in gaming.

We considered more number of participants in evaluating the Rubik's cube solution methods as compared to the other evaluations as the former comprises of only a single question of giving scores to each solution method. Whereas the other evaluation techniques have a large number of questions giving us more data points to analyse even with lesser number of participants as it has better chances of unfolding them more issues.

The users were encouraged to finish playing the entire game but they had to complete laying a full pipeline between the tank and a house at the minimum. For the \emph{baseline} measurements, the users performed this task without having to solve the Rubik's cube, i.e., they directly started the pipe laying task with all assets already present in the inventory. For recording measurements about the hybrid system, the full game including Rubik's cube solve was played. There was no time limit given and users were allowed to ask questions throughout the process whenever they had some doubts. All participants could complete the given task in a single attempt, except $2$ who needed another chance. The most common mistakes that were made included rotating the cube in a direction opposite to the one displayed and choosing incorrect pipes for elevated portion of the terrain. The subsequent sections discuss evaluations of the complete hybrid game and the Rubik's cube solve.

\section{Hybrid game evaluation}\label{gameMechEval}
In this section, we analyse the interaction mechanism and overall gameplay of the hybrid game along with its effect with respect to multiple measurements of usability, cognitive load and appeal to the user. 

\subsection{General questionnaire}\label{susEval}
We designed a questionnaire combining some questions from the SUS questionnaire \cite{brooke1996sus} and questions relevant to our game design with the objective of capturing the overall trend about how enjoyable, easy-to-use and involving the proposed gameplay is:

\begin{itemize}
    \item[-] (S1) I found the gaming experience enjoyable.
    \item[-] (S2) I did not find the gameplay interesting.
    \item[-] (S3) I could easily adapt to the gameplay without any technical support from the team.
    \item[-] (S4) I found the game and its components unnecessarily complex.
    \item[-] (S5) I think the Rubik's Cube component of the game blended well with the rest of the game.
    \item[-] (S6) I found the visual cues for solving Rubik's Cube more helpful than the written instructions for it.
    \item[-] (S7) I  think the game is quite lengthy and tends to make the player lose interest over time.
    \item[-] (S8) I feel this AR based tabletop game is more engaging than any other similar 2D tabletop game.
    \item[-] (S9) I think I would play the game frequently.
    \item[-] (S10) I think the game would be an interesting learning experience for most people.
\end{itemize}
A higher score for a question which is formed by a negative statement about the game (for example, question S2) is an indicator that the statement is not strong. It is trivial to infer that a higher score for a question which is formed by a positive statement about the game (for example, question S1) is desirable.
\begin{figure}[h!t]
    \centering
    \subfloat[S1]{\includegraphics[width=0.3\linewidth]{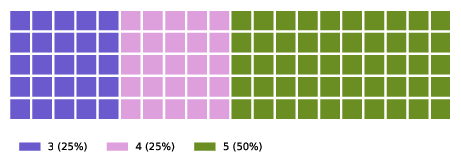}}~~
    \subfloat[S2]{\includegraphics[width=0.3\linewidth]{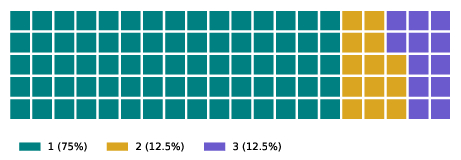}}~~
    \subfloat[S3]{\includegraphics[width=0.3\linewidth]{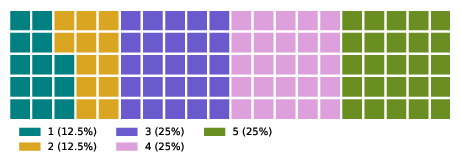}}~~
    
    \subfloat[S4]{\includegraphics[width=0.3\linewidth]{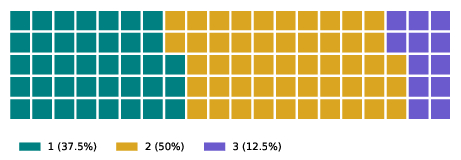}}
    \subfloat[S5]{\includegraphics[width=0.3\linewidth]{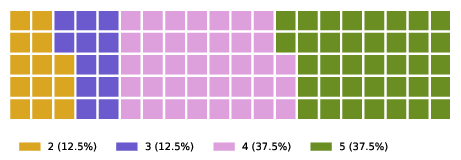}}~~
    \subfloat[S6]{\includegraphics[width=0.3\linewidth]{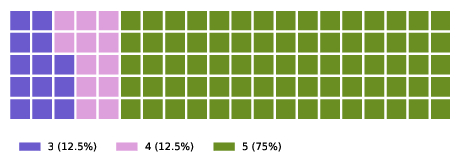}}~~
    
    \subfloat[S7]{\includegraphics[width=0.3\linewidth]{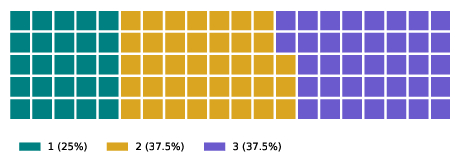}}~~
    \subfloat[S8]{\includegraphics[width=0.3\linewidth]{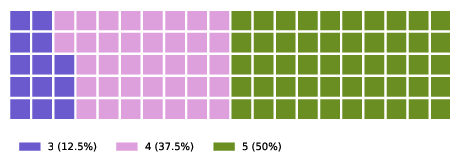}}
    \subfloat[S9]{\includegraphics[width=0.3\linewidth]{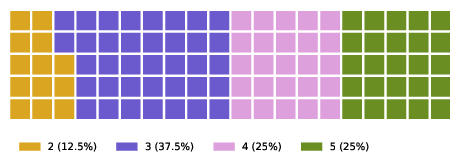}}~~
    
    \subfloat[S10]{\includegraphics[width=0.3\linewidth]{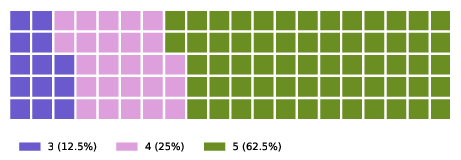}}
    
    \subfloat{\includegraphics[width=0.8\linewidth]{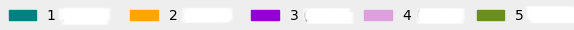}}
    \caption{Distribution of the general question scores for the hybrid game. The legends below all the plots shows the mapping between the colors and scores ($1$-$5$).}
    \label{figcomSUS}
\end{figure}
Figure~\ref{figcomSUS} depicts the distribution of scores given by the users for the individual questions. For e.g., $25\%$ of the users gave a score of $3$ and $4$ each, and $50\%$ of the users gave a score of $5$ for question S1 while evaluating the game. Note that the waffle charts show the percentage wise distribution of the scores given by users for a particular question. Each cell does not represent a user rather a cluster of cells filled with same color represents the percentage of users who gave the corresponding score. We use a subset of relevant questions for the \textit{baseline} measurements. Additional results from this questionnaire for the baseline and the hybrid game may be found in the supplementary PDF document.

The pipe laying task built around the cube solving challenge makes the overall gameplay interesting while not not making it too complex due to the addition of cube solving challenge to it (see scores for question S4 in Figure~\ref{figcomSUS}d), and can be considered as trade-off for a more interesting gameplay (see scores for question S2 in Figure~\ref{figcomSUS}b). Moreover, these two components blend very well with each other while not interfering with the performance in the individual tasks (see scores for question S5 in Figure~\ref{figcomSUS}e). The results also show that the game is potentially a good learning experience and ease-to-use can be further improved by evolving a more visually nuanced interface. 
Quantitatively, the lowest score among all other questions is reported on the question S3 ($2.4$ on the scale of $4$). This can be attributed to the lesser acclimatization time to the AR environment, as most of the users were not much used to AR.  The relatively lower score for S7 and S9 ($2.9$ and $2.6$ on the scale of $4$) indicates a need for a more nuanced interface to make the experience visually better. The questions S1 and S10 are among the top scorers (above $3$ on the scale of $4$) showing enhanced and richer user experience due to overall challenges present in the game. 




\subsection{NASA TLX questionnaire}\label{nasatlxEvalgm}
The NASA Task Load Index (TLX) evaluates the effect of an interface on an user in terms of physical, mental and temporal demands. It tries to quantify the cognitive load of an interface on the user by determining the fatigue caused by its usage. We choose the following standard NASA TLX questions for the user:

\begin{itemize}
    \item[-] (N1) Mental Demand: The game was easy to understand and play.
    \item[-] (N2) Physical Demand: The game was very strenuous to play.
    \item[-] (N3) Temporal Demand: The pace at which each task of the game proceeded was slow.
    \item[-] (N4) Overall Performance: The performance in the game is satisfactory.
    \item[-] (N5) Effort: It was easy to accomplish the level of performance.
    \item[-] (N6) Frustration Level: The gaming experience is relaxing and not stressful.
\end{itemize}

We calculate the load index for each user as per the method explained in~\cite{hart1988development}. A lower load index for a user shows that the user had a stress-free and enjoyable experience while using the interface. A lower mean load index for each question shows a positive trend with respect to that factor.
\begin{figure}[h!t]
    \centering
    \subfloat[N1]{\includegraphics[width=0.30\linewidth]{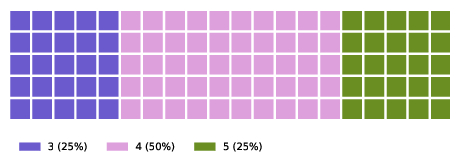}}~~
    \subfloat[N2]{\includegraphics[width=0.30\linewidth]{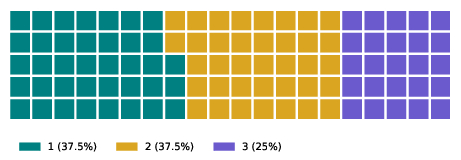}}~~
    \subfloat[N3]{\includegraphics[width=0.30\linewidth]{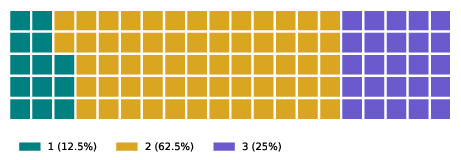}}~~    
    
    \subfloat[N4]{\includegraphics[width=0.30\linewidth]{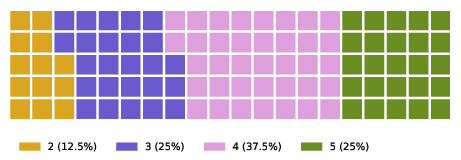}}~~
    \subfloat[N5]{\includegraphics[width=0.30\linewidth]{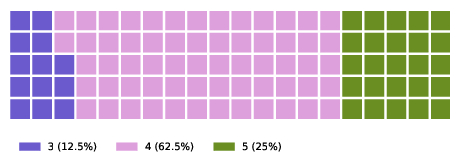}}~~    
    \subfloat[N6]{\includegraphics[width=0.30\linewidth]{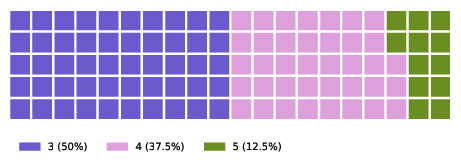}}     
    
    \subfloat{\includegraphics[width=0.8\linewidth]{pictures/legendsQue.png}}
    \caption{Distribution of NASA TLX question scores for the hybrid game. The legends below all plots shows the mapping between the colors and scores ($1$-$5$).}
    \label{figcomNTLX}
\end{figure}
The waffle charts in Figure~\ref{figcomNTLX} depict the distribution of scores (on a scale of $1$ to $5$) given by each user for the individual questions for the hybrid game. TLX scores are mapped on a scale of $100$ for analysis. Additional results from the NASA TLX questionnaire for the baseline and the hybrid game may be found in the supplementary PDF document.

The factor N3 turns out be the highest load factor ($42.5$) in the entire game. This shows there is scope of improvement for the interface to further optimize various operations for a smaller response time. We present detailed performance/timing parameter measurements later in Section~\ref{otherPar}. Our proof-of-concept implementation can be further optimized for speed. The factors N1, N4 and N5 ($20$, $25$ and $17.5$) are among the lowest load factors. This can be attributed to the assistance that we provide at each step of the game like the visual cues to help solve the cube or disallowing the placement of certain assets at some locations on the terrain. The cognitive load of the game is well within acceptable limits. This is inferred from the average load index of $32.1$ (as compared to our score of $26.2$) given in \cite{chu2020comparing} for the proposed AR based system which is claimed to be acceptable. 

\subsubsection{Statistical significance measurement}
The null hypothesis is: 

$H_0$: \emph{The workload for a hybrid game with both virtual and tangible elements is identical to the workload of a game with only virtual elements}. 

The load index of the baseline is $23.2$. A one-tailed t-test with our average load index of $26.2$ ($\sigma = 12.2$) gives us the p-value to be $0.255$. This is greater that the significance level alpha of $5$\%. Therefore, the t-test supports the null hypothesis implying that the difference between the load index of a hybrid game versus that of a non-hybrid game is not statistically significant. Therefore, adding the tangible challenge puzzle interface to the baseline game constitutes a viable and engaging game design with no significant increase in the workload. This answers our research question, \textbf{R3}.
\subsection{Presence questionnaire}\label{presenceEvalgm}
The presence questionnaire evaluates an interface from the point of view of the user in a more subjective manner. It brings out the trade off between the involvement and effort for adapting to an interface. It indicates how the user perceives the interface to be. This can be considered as a form of engagement with respect to how supportive an interface is for the user while performing some action. We do not evaluate the behavioural aspect of engagement, rather we evaluate its technical aspects in form of how much effective is the interface in elevating the sense of presence of the user. We use a relevant subset of the presence questionnaire given in~\cite{ntokas2015usability}. The presence questionnaire questions are categorized into at least one of the following criteria from each category:

Category I - CF = control factors, SF = sensory factors, DF = distraction factors, RF = realism factors, INV = involvement

Category II - RE = Realism, PA = Possibility to act, QI = Quality of Interface, PE = Possibility to Examine, SE = Self Evaluation of Performance
\begin{figure}[h!tb]
    \centering
    \subfloat[SF]{\includegraphics[width=0.30\linewidth]{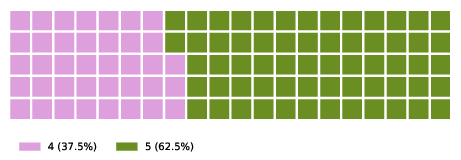}}~~~
    \subfloat[RF]{\includegraphics[width=0.30\linewidth]{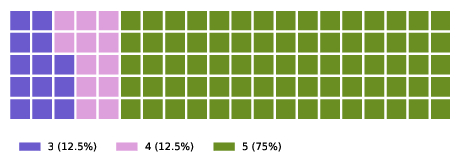}}~~~
    \subfloat[INV]{\includegraphics[width=0.30\linewidth]{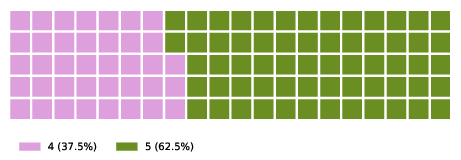}}~~~
    
    \subfloat[PE]{\includegraphics[width=0.30\linewidth]{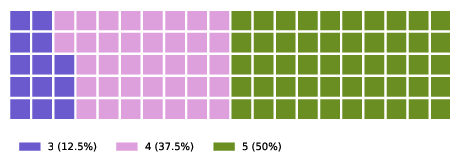}}~~~
    \subfloat[SE]{\includegraphics[width=0.30\linewidth]{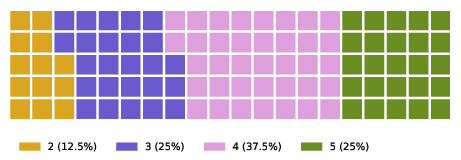}}~~~
    \subfloat[CF]{\includegraphics[width=0.30\linewidth]{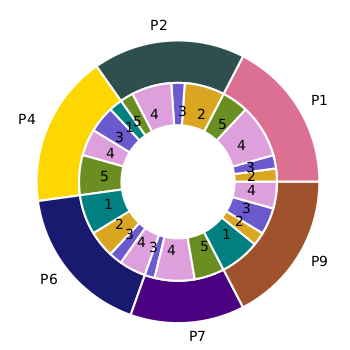}}~~~
    
    \subfloat[DF]{\includegraphics[width=0.30\linewidth]{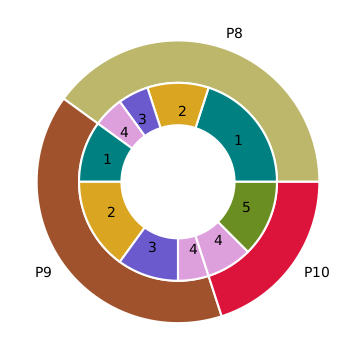}}~~~
    \subfloat[RE]{\includegraphics[width=0.30\linewidth]{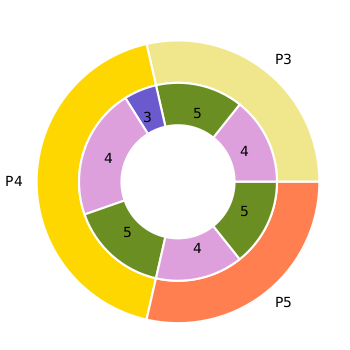}}~~~
    \subfloat[PA]{\includegraphics[width=0.30\linewidth]{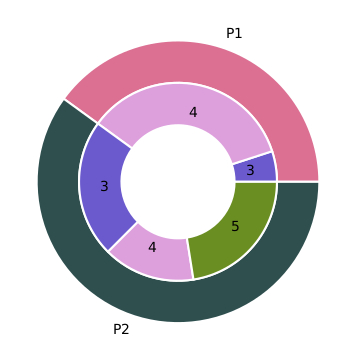}}~~~
    
    \subfloat[QI]{\includegraphics[width=0.30\linewidth]{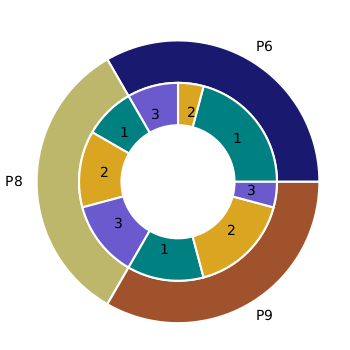}}
    
    \subfloat{\includegraphics[width=0.8\linewidth]{pictures/legendsQue.png}}
    
    \subfloat{\includegraphics[width=\linewidth]{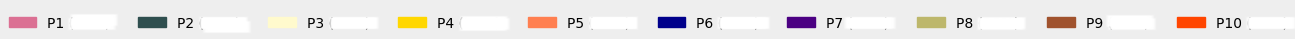}}

    \caption{Distribution of Presence question scores for the hybrid game. The legends below each waffle chart show the mapping between the colors and scores (in the range of $1$-$5$). The labels on top of every outer sector for each multi-level segmented pie chart shows the color and question mapping and the numbers written within its inner sectors show mapping between the colors and the scores (in the range of $1$-$5$). The second mapping is consistent with that in the waffle charts. The legends below all the plots show these mappings.}
    \label{figcomPres}
\end{figure}

The criteria for each question is specified along with the question. We choose the following presence questions for the user to answer:
\begin{itemize}
    \item[-] (P1) How much were you able to control events? (CF/PA)
    \item[-] (P2) How responsive was the environment to action that you initiated (or performed)? (CF/PA)
    \item[-] (P3) How much did the visual aspects of the environment involve you? (SF/RE)
    \item[-] (P4) How much did your experiences in the augmented environment seem consistent with your real-world experiences? (RF/CF/RE)
    \item[-] (P5) How involved were you in the augmented environment experience? (INV/RE)
    \item[-] (P6) How much delay did you experience between your actions and expected outcomes? (CF/QI)
    \item[-] (P7) How quickly did you adjust to the augmented environment experience? (CF/SE)
    \item[-] (P8) How much did the visual display quality interfere or distract you from performing tasks or required activities? (DF/QI)
    \item[-] (P9) How much did the control devices interfere with the performance of tasks or with other activities? (DF/CF/QI)
    \item[-] (P10) How well could you concentrate on tasks rather than on the mechanisms used to perform them? (DF/PE)
\end{itemize}
We calculate the mean and other statistical measures for each of the factors. The kind of desirable score whether lower or higher for a factor depends on the distribution of the questions that belong to it, i.e., the ratio of questions for which a negative feedback is desirable to that of questions for which a positive feedback is desirable. Figure~\ref{figcomPres} depicts the distribution of scores given by each user for the individual questions of each factor in both categories for the hybrid game. Responses for factors that belong to multiple questions are shown as a multi-level segmented pie chart. For example in Figure \ref{figcomPres}g, it can be seen that the plot has $3$ outer sectors corresponding to each questions corresponding to the factor. Considering the sector for question P10, it can be seen that the users gave scores of $4$ and $5$ which constitute the inner sectors. The arc length of these inner sectors represent the percentage of users who gave a particular score. Factors that belong to a single question are shown as waffle charts as earlier. Detailed quantitative results for the presence questionnaire for baseline and hybrid game are given in the supplementary PDF document.

The control factors (CF) have $6$ questions with $4$ for positive feedback and $2$ for negative feedback, so our score of $3.3$ can be considered as a good score. According to the distraction factors (DF) question distribution, a score below average ($2.5$) and a bit above the lowest ($1$) should be considered as a good score. In our case ($2.9$), that is not true which can be partially due to the combination of both the components which requires the user to interleave these two tasks. Moreover, slower response to the change of the face in case of augmented cube at times can also be a major reason behind this. But we can observe that in spite of this, the score for the question P10 (see Figures~\ref{figcomPres}d and \ref{figcomPres}g) is higher which indicates that the user is not completely distracted while playing the game. The high scores (all above $4$) in sensory factors (SF), realism factor (RF), involvement (INV), realism (RE), and possibility to examine (PE) are major indicators of the effectiveness of the AR environment in general. Also, the 3D nature of the game helps the user to get an illusion of doing the tasks in real world, albeit at a small scale. According to the possibility to act (PA) and quality of interface (QI) question distributions, our scores of $3.9$ and $1.8$ respectively, are fine. The slight drop in the PA score and $0.3$ difference in the QI score for the hybrid set-up from the baseline ($4.2$ and $1.5$) may be due to the interleaving between two tasks and perceived delay in actions. The scores of the self evaluation (SE) of environment deviates in case of the baseline and the hybrid set-up ($4.2$ and $3.8$). This can be due to the increased complexity of the hybrid set-up along with the lesser acclimatization time given to the volunteers to get used to the AR environment. 

\subsubsection{Statistical significance measurement}
The null hypotheses for category I and category II are:

$H_{01}$: \emph{The involvement with the interface for the hybrid game with virtual and tangible elements, and the baseline with only virtual elements is the same.}

$H_{02}$: \emph{The effort required to interact with the hybrid game with virtual and tangible elements, and the baseline with only virtual elements is the same.}

The baseline presence scores in the two categories are $39.6$ and $34$ respectively. A one-tailed t-test with our scores of $42.1$ ($\sigma = 4.2$) and $34.8$ ($\sigma = 2.6$) gives us the p-value to be $0.065$ and $0.188$ for the two categories respectively. Both the p-values are greater than the significance level alpha of $5$\%, implying that the null hypothesis is supported in both cases. Thus, we can conclude that while interacting with a hybrid game does not require additional effort over the baseline, it also does not entail any significant improvement in user involvement. This points to the fact the hybrid interface can be further improved. This analysis provides valuable insights towards answering our research question \textbf{R1}.

From the above discussions, it can be inferred that the hybrid gameplay makes the baseline game (a pure virtual game) more interesting while not making it too much complex and stressful. That is, AR based hybrid systems are indeed usable and engaging to the users.
\section{Cube solution evaluation}\label{rubikCubeEval}
In this section, we analyse the effectiveness of the assistance provided by our interface in form of visual  augmentations for solving the Rubik's Cube with respect to the solution method as well as its impact on the game design. We attempt to answer \textbf{R2} in this section by analyzing the outcomes of the user responses. Additional plots for the evaluation from TLX and presence questionnaires can be found in the supplementary PDF document.
\subsection{NASA TLX questionnaire}\label{nasatlxEvalrc}
The same set of questions as given in section \ref{nasatlxEvalgm} are used and similar post processing method is followed to evaluate the augmented cube solve interface.
\begin{table}[h!t]
    \centering
    \begin{minipage}{0.5\textwidth}
        \centering
        \begin{tabular}{|c|c|c|c|c|c|c|}
        \hline
        & N1 & N2 & N3 & N4 & N5 & N6 \\ \hline
        $\mu$ & 8 & 32 & 36 & 12 & 20 & 16 \\ \hline
        $\sigma$ & 10.9 & 26.8 & 21.9 & 10.9 & 14.1 & 26.1 \\ \hline
        \end{tabular}
        \caption{NASA TLX evaluation results for augmented cube.}
        \label{userEvalaugNTLX}
    \end{minipage}~%
\end{table}
\begin{table}[h!t]
    \begin{minipage}{0.5\textwidth}
        \centering
        \begin{tabular}{|c|c|c|c|c|}
        \hline
        $\mu$ & Median & Min. & Max. & $\sigma$ \\ \hline
        22.7 & 20 & 8 & 53.3 & 17.9 \\ \hline
        \end{tabular}
        \caption{Overall TLX load indices for augmented cube.}
        \label{userEvalscoreNTLXrc}
    \end{minipage}    
\end{table}

From table \ref{userEvalaugNTLX}, we see that factor N3 turns out to have the highest load factor, following a similar trend as for the hybrid game. This shows the scope of improvement for the cube solving mechanism with respect to the response time, see Section \ref{otherPar} for more details. 

In \cite{chen2018investigation}, Chen et al. compare the psychological effect of learning to solve a Rubik's cube using direct instructions and minimal guidance methods. The direct method where the instructions to solve the problem are given to a participant by various media can be considered as baseline for our method of using visual cues. Considering their reported average load index of $57.3$, our index value of $22.7$ is well within acceptable range.
\subsection{Presence questionnaire}\label{presenceEvalrc}
The same set of questions as given in section \ref{presenceEvalgm} are used and similar post processing method is followed to evaluate the augmented cube solve interface.
\begin{table}[h!t]
    \centering
    \begin{minipage}{0.5\textwidth}
        \centering
        \begin{tabular}{|c|c|c|c|c|c|c|c|}
        \hline
        Category I & CF & SF & DF & RF & INV \\ \hline
        $\mu$ & 3.7 & 4.4 & 2.5 & 4.6 & 4.6 \\ \hline
        $\sigma$ & 0.9 & 0 & 1.4 & 0 & 0 \\ \hline \hline
        Category II& RE & PA & QI & PE & SE \\ \hline
        $\mu$ & 4.5 & 4.1 & 1.9 & 4.4 & 4.6 \\ \hline
        $\sigma$ & 0.1 & 0.1 & 0.6 & 0 & 0 \\ \hline
        \end{tabular}
        \caption{Presence I \& II evaluation results for augmented cube.}
        \label{userEvalaugPres}
    \end{minipage}~~%
\end{table}
\begin{table}[h!t]
    \begin{minipage}{0.5\textwidth}
        \centering
        \begin{tabular}{|c|c|c||c|c|}
        \hline
        &\multicolumn{2}{|c||}{Category I}&\multicolumn{2}{c|}{Category II}\\ \hline
        & Total & Scaled & Total & Scaled \\ \hline
        $\mu$ & 43 & 3.6 & 36.6 & 3.7 \\ \hline
        Min. & 36 & 1.4 & 31 & 1.4\\ \hline
        Max. & 49 & 4.6 & 41 & 4.6\\ \hline
        $\sigma$ & 5.3 & 1.2 & 3.8 & 1\\ \hline
        \end{tabular}
        \caption{Overall Presence I and II scores for augmented cube.}
        \label{userEvalscorePresrc}
    \end{minipage}
\end{table}

We observe a $0.3$ difference between the scores of the augmented cube component ($3.7$) and the hybrid game ($3.3$) with respect to control factor (CF). This can be attributed to the fact that the cube gives the user a physical entity to manipulate as compared to using a mobile device to control events. The presence of a physical entity gives an impression of more freedom to users pointing towards potential advantages of combination of tactile feedback and visual feedback. Table \ref{userEvalscorePresrc} shows that the augmented cube component follows a similar trend as that of the hybrid game indicating it to be a positive addition. 
\subsection{Rubik's Cube solution method comparison} \label{rubikComp}
We further evaluate the effectiveness of the visual cues in solving the Rubik's Cube as given in our game to more commonly available set of written instructions, as well as to solving it without any help.

\begin{figure}[h!t]
    \centering
    \subfloat[Written instructions]{\includegraphics[width=0.45\linewidth]{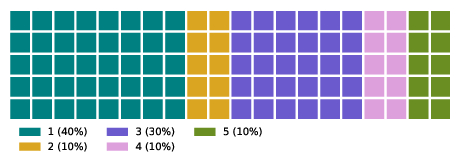}}~
    \subfloat[Visual Cues]{\includegraphics[width=0.45\linewidth]{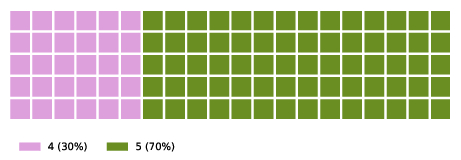}}
    
    \subfloat{\includegraphics[width=0.8\linewidth]{pictures/legendsQue.png}}
    \caption{Evaluation of the methods for solving the Rubik's Cube. The legends below each plot shows the mapping between the colors and scores (in a range of $1$-$5$).}
    \label{figrubikComp}
\end{figure}

\begin{table}[h!t]
    \centering
    \begin{minipage}{0.4\textwidth}
        \centering
        \begin{tabular}{|c|c|c|c|}
        \hline
        & None & Written & Our method \\ \hline
        $\mu$ & 1.4 & 2.4 & 4.7 \\ \hline
        Median & 1 & 2.5 & 5 \\ \hline
        Min. & 1 & 1 & 4 \\ \hline
        Max. & 3 & 5 & 5 \\ \hline
        $\sigma$ & 0.7 & 1.4 & 0.5 \\ \hline
        \end{tabular}
        \caption{Overall user evaluation of methods to solve the Rubik's Cube.}
        \label{userEvalrubik}
    \end{minipage}\hspace{20px}%
    \begin{minipage}{0.5\textwidth}
        \centering
        \begin{tabular}{|c|c|c|c|c|}
        \hline
        & Reaching targets &\multicolumn{2}{|c|}{Cube solve}& Pipe Laying\\ \hline
        & & Phase I & Phase II &\\ \hline
        Time & 30 & 250 & 265 & 270 \\ \hline
        Steps & 2 & 21 & 30 & 50 \\ \hline
        \end{tabular}
        \caption{Average time(in seconds) and number of steps for completing tasks in the game.}
        \label{compTimes}
    \end{minipage}
\end{table}

Figure~\ref{figrubikComp} depicts the distribution of scores given by each user for the written instruction based and the visual cue based methods to solve the Rubik's Cube. The preference for the visual cues is apparent.
Table~\ref{userEvalrubik} shows the mean ($\mu$), median, minimum, maximum and standard deviation ($\sigma$) of the score (on a scale of $5$) for the methods used to solve the Rubik's Cube. Here we also include a method where no instructions were available to the users on how to solve the Rubik's Cube. An important observation can be made during the user evaluation sessions -- the written instruction method is easier for someone familiar with the movement notation or the algorithm to solve the cube, however, for a novice user understanding the written instructions are very tiresome and the visual cues help them to understand the movements better.

\section{Timing performance parameters}\label{otherPar}
The performance of an interface can also be defined in terms of the average time and number of steps to complete its sub-tasks. In case of AR based interfaces, the stability of augmentations is also important. In our case, the augmentations on the markers are quite stable at larger distances, although there can be small drifts in the positions of assets due to readjustment of 3D positions by ARCore. Also, prolonged usage of ARCore on a mobile device, in particular with rapidly changing marker images (as is the case of the Rubik's Cube), causes it to heat up which causes throttling of the device processor speed and results in some lag between various actions.

Table \ref{compTimes} shows the average completion times and average number of steps (taps on screen or rotations to cube) for major tasks of the gameplay averaged over $5$ runs of the game. In the table, by "reaching targets" we denote the movement of the virtual player between two points on the terrain to reach static or dynamic assets, and "pipe laying" denotes the actions of the user to complete the goal after acquiring required assets. For cube solve, the complete solution is split into two phases. In phase I, the bottom face and the bottom layer of all other faces except the top face is solved. In phase II, the remainder of the cube is solved. See the supplementary PDF for more details.
The phase I cube solve includes an average time of $84$ seconds and $6$ steps for the one-time step of scanning all faces of the cube initially. It takes around $35$ seconds on an average to recognise markers on all cells of a face. This detection task is faster in the beginning and degrades with time due to reasons stated above. On an average, it takes around $815$ seconds and $103$ actions to complete a round of the entire game.

\section{Conclusion and Discussion}\label{disc}
We design and present an interactive, AR based hybrid game called \textit{Cube2Pipes}. The gameplay is a unique combination of a virtual tabletop game and a tangible challenge puzzle interface of a $2\times2$ Rubik's Cube. A thorough user evaluation of the complete game is done using well established methods. 
The use of the tangible component in the hybrid game makes it interesting as per general questionnaire scores. The workload on the user does not increase much with the extra level of challenge by adding Rubik's Cube solution to the pipe laying problem as per NASA TLX score. The addition of solving Rubik's Cube to the game does not add to the effort of playing the game while maintaining a degree of involvement with the interface as per the Presence evaluation. Also, the visual augmented cues based method of solving the Rubik's Cube independently has a good Presence score, which is congruent with the results of solution method comparison study.
We are thus able to conclusively infer that such a hybrid game is convincing and appealing to the user. As an add-on advantage of this gaming mechanism, the user is also able to understand the spatial moves required to solve the Rubik's Cube in an organic way as demonstrated by Rubik's cube solution method comparison scores.

A limitation of our game is the quality of graphical elements. The visual design of the assets and the movement animation of the character in the game can be substantially improved. Though we observe that in spite of this, the users found the gameplay interesting and engrossing. A major implementation challenge is to resolve the various lags and delays that occur due to a prolonged usage of the AR platform or quickly changing image markers. We want to extend the challenge mechanism to a $3\times3$ Rubik's Cube, however, the degrees of freedom of the solution space for the cube makes it a very tricky problem to solve.

In a more general sense, the results presented show the potential of any such interactive hybrid system in providing an enjoyable and interesting experience to users. These results can generalize to the entire class of all such hybrid systems as our baseline and hybrid game comparisons are independent of the tasks and revolves more around the impact of the blend of virtual and tangible components. It also shows the immense potential that AR platforms offer in facilitating learning. With the help of visual and tangible AR interactions, users can potentially smoothly understand the solutions to other similar complex yet interesting puzzle-like problems belonging to the class of the game presented in the paper. Based upon the outcomes of our user studies, we postulate that for younger audiences, assumed to have less prior knowledge, an AR based system can actually be used for showing the process involved in various activities ranging from how water is supplied to their homes to constructing roads to assembling an airplane in form of a 3D interface, showing which is otherwise infeasible. Of course, the issue of getting users more comfortable with using AR based interfaces and filling the gap between a near seamless experience coherent with the real world without wearing out the user too much needs to be further addressed. Our studies also point towards potential directions for improving the interface.
Major take away from the findings of the user studies presented in the paper are-- using AR based visual cues for assisting users to understand tricky challenges and puzzles provides an interesting future direction to explore, and design choices for hybrid gameplays in non-immersive environments like AR should consider more nuanced interfaces to further involve users. Overall, from the results of the user studies it can be said that probing more into such hybrid designs can potentially make AR based learning applications or games richer in experience and more interesting.
\bibliographystyle{abbrv-doi}
\bibliography{00ref}
\end{document}